# Time dependent FTIR spectra of mineral waters after contact with air

[1*]Kondyurin A., [2]Sadikov Ya.

[1]School of Physics, University of Sydney, Sydney, Australia

[2]Perm State Technical University, 614600 Perm, Russia

## Abstract

FTIR spectra of mineral waters of Slavyanovskaya, Aqua Montana, Bad Harzburger and Christinen with time from first contact of water with open air were analysed. The kinetic of spectral changes of Slavyanovskaya mineral water in the regions of stretch, deformation and intermolecular vibrations was measured. The spectral changes do not correlate with chemical contamination of mineral water and degassing process. The observed spectral changes could be due to different structure of mineral water in liquid state, which is destroyed after air contact. The observed spectral behaviour of Slavyanovskaya is correlated with the catalytic activity of mineral water, which was saved without contact with air. The characteristic time of spectral dependence (669 seconds) is close to the characteristic time of catalytic activity loss (600 seconds) of mineral water at air contact. The spectra results support the medical studies that show the activity of mineral water near spring, and the loosing activity of water after long time storing.

Keywords: water, FTIR spectra, structure, health resort.

## Introduction

Water is unusual liquid. A wide number of theoretical models of water structure in liquid state are developed [1-5]. But up to now, no one developed a theory, which explains completely all water behaviour and properties in liquid state. One of the modern theories is based on cluster structure of associated water molecules in liquid state [6-8]. The cluster structure of water is formed by dense network of hydrogen bonds, which support an ordered orientation of water molecules at long distance [9]. Today, this view on water structure is mostly accepted. However, the cluster structure of water was not experimentally observed.

One of the interesting forms of water is a natural mineral water. The mineral water is a natural therapeutical remedy for a wide number of illnesses and powerful prophylactic medicine. It was observed some times, that maximal therapeutical effect of mineral water is achieved near water spring, when the time between water release from underground and drinking is not longer then 10-15 minutes. These observations were supported by measurement of mineral water properties from different natural springs and with different methods of mineral water analysis. The first observation of specific catalytic activity of natural mineral water was found in the description of health resort Vichy (France) in 1907 [10]. Catalytic activity was determined by the ability of mineral water to destroy hydrogen peroxide added to mineral water immediately after receiving of water from underground spring. The catalytic activity of mineral water to destroy hydrogen peroxide disappears with time at saving in bottles [11]. The inactivation of water was explained by the air contact.

To exclude the air contact, a technology of getting, storing and transportation of active mineral water was developed and described in [12-14]. At this technology the mineral water is taken from underground without air contact, where the initial concentration of carbon dioxide remains. The special containers are designed and used for transportation of active mineral water.

The analysis showed, that the salt contamination, carbon dioxide concentration and catalytic activity remain in mineral water stored in these containers during 6 month [15]. But the catalytic activity of water stored in bottles or on open air is lost. The time, when the catalytic activity is lost at open air, is about 10 minutes after release from underground spring [16]. The distillate water does not have the catalytic activity. Inserting of carbon dioxide to water does not lead to appearance of catalytic activity of distillate water and does not reverse activity in previously inactivated mineral water. The catalytic activity and mineralisation of natural mineral waters from different sources in Russia, Georgia, Armenia and Czech were measured [14]. The correlation of catalytic activity and salt concentration or carbon dioxide concentration was not found for mineral water from different underground springs.

The long-term clinical investigations of the patients in Perm hospital (Ural, Russia, about 2000 km far away from natural springs of mineral water) and the patients in Zheleznovodsk hospital (Caucasus region, Russia, near natural source of mineral water) were carried out [10]. A gastric pH of patients with pathologies of stomach-intestines system was monitored before and after the patients drunk the mineral water of Zhelesnovodsk springs (Russia). The effect of the active mineral water from ground source, the active water from container and the inactive water from bottles was investigated. The using of active mineral water immediately from source and from special container gives the normal value of gastric pH of patients during a longer time then the using of inactive mineral water from bottles. The patients with anomalies of chronicle gastritis with low and high secretions, stomach ulcer, with liver and pancreas problems were treated by active and inactive mineral water. The using of active mineral water from containers decreases the numbers of ill cases, ill cases at finger tests and dyspepsia cases in comparison with using of inactive mineral water from bottles. All these studies showed a specific effect of active mineral water on patients.

The active and inactive mineral water were analysed by Raman spectra [17]. The spectra of inactive mineral water has maximums of vibration state density at 183, 512 and 798 $cm^{-1}$, the spectra of active mineral water has peaks at 180, 500, 617 and 914 $cm^{-1}$. Additional peaks were interpreted as intermolecular vibrations of specific intermolecular structure of water in liquid state.

Despite on interest to mineral water and therapeutic effects, the properties of active mineral water were not investigated. In present work, the FTIR spectra of active mineral water are analysed with time after removing of active mineral water from the container.

**Experiment**

Mineral water of „Slavyanovskaya" source (Zheleznovodsk, Russia) was extracted without air contact with equipment described in [14] and transported in special containers described in [13] under carbon dioxide atmosphere of 0.4 MPa pressure. Test of hydrogen peroxide decomposition showed, that the catalytic ability of mineral water remains as in fresh mineral water.

Mineral waters from bottles Aqua Montana (Acque Minerali val Menaggio S.p.A., Italy), Bad Harzburger (Bad Harzburger Mineralbrunnen GmbH, Germany) and Christinen (Teotoburger Mineralbrunnen GmbH, Germany) were analysed for comparison. Spectra of distillate water and inactivated mineral water stored on air during one day were recorded for comparison. Chemical composition of mineral waters from [18] and product information is presented in Table.

FTIR transmission spectra were recorded on Bruker IFS-66s and Nicolet Magna 750 spectrometers with polyethylene cell (50 mkm low density polyethylene films between KBr pellet). The resolution was 2 $cm^{-1}$, the number of scans was 10. The cell in the spectrometer was filled by water from the container or from bottle and first scan of spectra was started in 30 seconds after first contact of water with air. Time of spectra registration did not exceed 10 seconds for Bruker and 20 second for Nicolet. Minimal time between the spectra recording was 30 seconds.

**Results**

Spectra of distillate and inactive mineral water contain wide band of stretch symmetric and asymmetric vibrations in region of 3600-3100 $cm^{-1}$ and band of deformation vibration in the region of 1650 $cm^{-1}$. The FTIR spectra of different waters are shown on Fig.1. In the region of 2149 $cm^{-1}$ an overtone of vibration is observed and in region of 400-600 $cm^{-1}$ some bands of intermolecular vibrations are observed [19]. In spectra of mineral waters a line of carbon dioxide is observed in region of 2337 $cm^{-1}$. Besides that, some strong bands of polyethylene cell are observed in regions of 3000-2700, 1500-1400 and 750-700 $cm^{-1}$.

The band of stretch vibration is a sum of overlapped separate lines corresponding to water molecules connected by different hydrogen bonds. The fitting of integral band with individual lines is questionable, because the band is broad and not structurized. The same complex shape of band is observed in the region of deformational vibrations and in the region of intermolecular vibrations. Such spectra are observed for distillate water and for mineral waters of Aqua Montana, Bad Harzburger and Christinen.

In comparison with spectra of distillate water and mineral water from bottles, in spectra of Slavyanovskaya mineral water the narrow separate peaks on the wide band of stretch vibrations are observed at 3684, 3669, 3645, 3625 and 3607 $cm^{-1}$. In the region of deformation vibrations the band is wide and two separate peaks at 1681 and 1645 $cm^{-1}$ are observed. In the region of intermolecular vibrations some separate peaks are observed at 620, 590, 544, 505 and 478 $cm^{-1}$.

Initial mineral waters and mineral waters in closed volumes contain soluble carbon dioxide. At first contact with air the carbon dioxide evaporates from water with formation of bubbles.

During degassing process the spectra lines of water vibrations have a lower intensity than for initial water due to bubbles formation. After the end of the degassing process the intensity of water lines return to initial values. The intensity of carbon dioxide line follows a decrease of its concentration in water to zero. For example the FTIR spectra of Christinen mineral water with time after opening of bottle are presented at Fig.2. After the finish of the degassing process observed by intensity of $CO_2$ vibration line at 2337 cm$^{-1}$ the spectra of mineral water does not change. The same picture is observed in spectra of Bad Harzburger and Aqua Montana mineral waters. For quantitative observation a dependence of spectra line intensities of stretch, deformation and $CO_2$ vibrations are presented in Fig.3 for Aqua Montana mineral water. Changes of line position and form of bands in spectra of this mineral water with time were not observed.

The FTIR spectra of Slavyanovskaya mineral water with time starting from first contact with air are presented in Fig.4 (a, b, c). The changes of spectra at first time (at degassing period) can be connected with bubbles formation and movement, that can cause a deformation of spectra. But after the degassing process the spectra of water in regions of stretch, deformation, and intermolecular vibrations continues to change. In the region of stretch and deformation vibrations the intensity of all lines increases. The intensity of 3373 cm$^{-1}$ increases. This line is observed in spectra of distillate water. The intensity of 3650, 3620 and 3600 cm$^{-1}$ lines increases too, but the relation of intensity of these lines to intensity of 3373 cm$^{-1}$ line decreases with time. In region of deformation vibrations the relation of intensity at 1681 cm$^{-1}$ line to the intensity at 1645 cm$^{-1}$ decreases. In the region of intermolecular vibrations the complex changes of spectra is observed with time. Especially, a very complex character of spectra transformations in this region is observed at first time after air contact. The line at 620 cm$^{-1}$ shifts to 629 cm$^{-1}$, shoulder at 595 cm$^{-1}$ appearances, line at 544 cm$^{-1}$ disappearances, line at 505 cm$^{-1}$ shifts to 515 cm$^{-1}$, shoulder at 478 cm$^{-1}$ growth. However, the spectra behaviour in intermolecular vibration region has not a good coincidence at spectra repetition with the same conditions. We suppose, that it can be connected with procedure of filling of cell by water from container and the conditions of first air contact.

The dependencies of spectral intensities of water vibrations and its relations in spectra of Slavyanovskaya mineral water are presented in Fig. 5, 6 and 7. The peak intensities of spectra were used in original values without deconvolution of spectra. The increase of stretch and deformation lines intensity is observed after the degassing process. Analysis of normalised intensities showed, that transformation of spectra in stretch and in deformation vibrations regions conducts by exponential law with characteristic times of 476 and 667 seconds accordingly. The degassing process conducts with characteristic time of 147 seconds. The difference of processes is clearly observed.

The behaviour of the intensity of intermolecular vibration lines is more complicate and does not correspond to the behaviour of intramolecular vibration lines of water molecule and the degassing process. The intensity of 505 cm$^{-1}$ lines decreases at first, does not change at degassing process and then it increases slowly. The intensity of 478 cm$^{-1}$ line sharply increases at first time, then it stops at degassing process and then it decreases slowly. Intensity of 620 cm$^{-1}$ line has more complex behaviour.

**Discussion**

The different behaviour of FTIR spectra of Slavyanovskaya from other mineral waters cannot be explained by different salt contamination. Concentration of salts is very low for significant changes of vibration spectra of all samples of mineral waters. Influence of ions on water spectra is observed at higher concentration of salts [20, 21]. The analysed mineral waters do not have a great difference in salt contamination, that can lead to significant spectral difference. The spectra do not correlate with the salt contamination and salt concentration observed in investigated samples. The low difference of pH in different mineral waters cannot lead significant spectral difference. Concentration of carbon dioxide is low in comparison with a limit of solubility of $CO_2$ in water. After the degassing process the spectral changes continue in the spectra of Slavyanovskaya mineral water. Therefore, the presence of carbon dioxide cannot cause the spectral differences of mineral waters. The spectra of water samples for all sources of mineral water do not show a presence of unexpected organic and inorganic substances in all analysed mineral water. Therefore, the chemical contamination cannot cause the difference in spectral behaviour of the mineral water samples. The spectral difference can be explained only by different structure of water in liquid state.

Smooth shape of stretch and deformational vibrations lines of distillate water corresponds to a homogeneous wide distribution of force constants and geometry of water molecules. Formation of hydrogen bonds changes the intramolecular force constants and geometry of molecules. The homogeneous distribution of intramolecular force constants and geometry of water molecules corresponds to a homogeneous distribution of energy and intermolecular force constants of the hydrogen bonds. Such distribution corresponds to statistical model of liquid, when a thermal movement of molecules disorders the cluster structure of water molecules and its intermolecular interactions.

The presence of narrow peaks in stretch vibration region of spectra of Slavyanovskaya supports, that some molecules have narrow distribution of force constants and geometry, which corresponds to narrow distribution of hydrogen bonds energy. That means, this water consists of clusters of molecules with similar force constants and geometry. Therefore, these molecules are bonded with similar hydrogen bonds and intermolecular distances as in a crystal.

In the stretch region of spectra the narrow peaks are observed in high frequency region of band, that corresponds to water molecules connected by weaker hydrogen bonds than other water molecules. It corresponds to theory of condensed state, when the molecules are densely packed to minimise the free spaces but the intermolecular interaction energy is not minimised. In this case, some of the water molecules cannot take part in formation of hydrogen bonds network or these molecules take part in weak hydrogen bonds. The structure of such formations of intermolecular interaction is usual for crystalline structure, when the packing of molecule is denser then in liquids. That means, Slavyanovskaya mineral water contains the associates or clusters of water molecules with dense packing and structure. Following dense packing requirements, such structure should be similar to the structure of crystalline water like ice.

Indirect confirmation of unusual structure of mineral waters can be supported the geological structure of Caucasus region. The water in Zhelesnovodsk sources has come as rain in Caucasus mountains about 10000 years ago. After slow flow at high pressure and high temperature without air the water can be packed in denser structure, then usual water. Then at normal pressure a structure of water transforms to usual less dense structure.

Other reasons of special structure of water can be connected with solution of air oxygen in water. At storage of mineral water in container without air oxygen during long time (2 and more years) the catalytic ability of water does not change. But at short time after getting of water from container or from ground source the catalytic activity of water drops down. Perhaps, the oxygen penetrates into water and destroys the water structure. The mechanism of destruction can be based on formation of hydroxide ions.

The structure transformation is sufficiently complex. The spectra of water comes to spectra of distillate water with time, that corresponds to formation of usual structure of water. The spectra of intermolecular vibration has complex behaviour, that corresponds to some stages of water structure transformation. Characteristic times of the spectra transformation in the region of stretch vibration (476 seconds) and in the region of deformation vibrations (667 seconds) are closed to the characteristic time of disappearance of catalytic activity of mineral water (about 600 seconds). This shows a connection between catalytic activity and spectral data of mineral water.

The presented results of spectra transformations and specific catalytic activity of mineral water cannot be explained with known and improved models of liquids. A new theory of water has to be developed.

**Conclusions**

Thus, the time depended FTIR spectra of different mineral waters were recorded. The catalytic active mineral water Slavyanovskaya has unusual FTIR spectra and complex spectral behaviour after contact with air. The spectral changes of active mineral water can be explained by a special structure of water in liquid state, which becomes destroyed after contact with air. The characteristic time of spectral changes is close to the characteristic time of the catalytic activity loss of mineral water.

Table. Contamination and properties of mineral waters.

| Parameter | Units | Slavyanovskaya | Aqua Montana | Bad Harzburger | Christinen |
|---|---|---|---|---|---|
| Integral salt concentration | g/l | 3,60 | 0,193 g/l | 0,227 g/l | no data |
| $[Na^+]$ | mol/mol | 5,52 $10^{-4}$ | 7,04 $10^{-7}$ | 6,26 $10^{-6}$ | 2,90 $10^{-4}$ |
| $[K^+]$ | mol/mol | 1,58 $10^{-5}$ | 9,22 $10^{-8}$ | no data | no data |
| $[Mg^+]$ | mol/mol | 3,45 $10^{-5}$ | 1,66 $10^{-6}$ | 6,00 $10^{-6}$ | no data |
| $[Ca^{+2}]$ | mol/mol | 1,19 $10^{-4}$ | 1,83 $10^{-5}$ | 1,71 $10^{-5}$ | 2,47 $10^{-5}$ |
| $[Sr^{+2}]$ | mol/mol | 5,34 $10^{-7}$ | no data | no data | no data |
| $[Mn^{+2}]$ | mol/mol | 3,74 $10^{-8}$ | no data | no data | no data |
| $[Fe^{+3}]$ | mol/mol | 2,05 $10^{-6}$ | no data | no data | no data |
| $[Zn^{+2}]$ | mol/mol | 2,23 $10^{-8}$ | no data | no data | no data |
| $[Cu^{+2}]$ | mol/mol | 2,55 $10^{-9}$ | no data | no data | no data |
| $[Co^{+2}]$ | mol/mol | 3,04 $10^{-10}$ | no data | no data | no data |
| $[SiO_2]$ | mol/mol | no data | 1,29 $10^{-6}$ | no data | no data |
| $[F^-]$ | mol/mol | 1,42 $10^{-6}$ | no data | no data | no data |
| $[Cl^-]$ | mol/mol | 1,46 $10^{-4}$ | 4,56 $10^{-7}$ | 5,07 $10^{-6}$ | 1,54 $10^{-4}$ |
| $[Br^-]$ | mol/mol | 4,05 $10^{-7}$ | no data | no data | no data |
| $[I^-]$ | mol/mol | 7,09 $10^{-8}$ | no data | no data | no data |
| $[SO_4^{-2}]$ | mol/mol | 1,66 $10^{-4}$ | 1,61 $10^{-6}$ | 8,99 $10^{-6}$ | 2,32 $10^{-5}$ |
| $[HCO_3^-]$ | mol/mol | 3,88 $10^{-4}$ | 6,42 $10^{-5}$ | 2,89 $10^{-5}$ | no data |
| $[HPO_4^{-2}]$ | mol/mol | 3,18 $10^{-9}$ | no data | no data | no data |
| $[NO_3^-]$ | mol/mol | no data | 2,18 $10^{-6}$ | no data | < 8,70 $10^{-8}$ |
| $[H_2BO_3]$ | mol/mol | 2,54 $10^{-6}$ | no data | no data | no data |
| $[H_2SiO_3]$ | mol/mol | 8,54 $10^{-6}$ | no data | no data | no data |
| $[CO_2]$, free | mol/mol | 5,27 $10^{-4}$ | 2,74 $10^{-6}$ | no data | no data |
| pH | - | 6,43 | 7,7 | no data | no data |

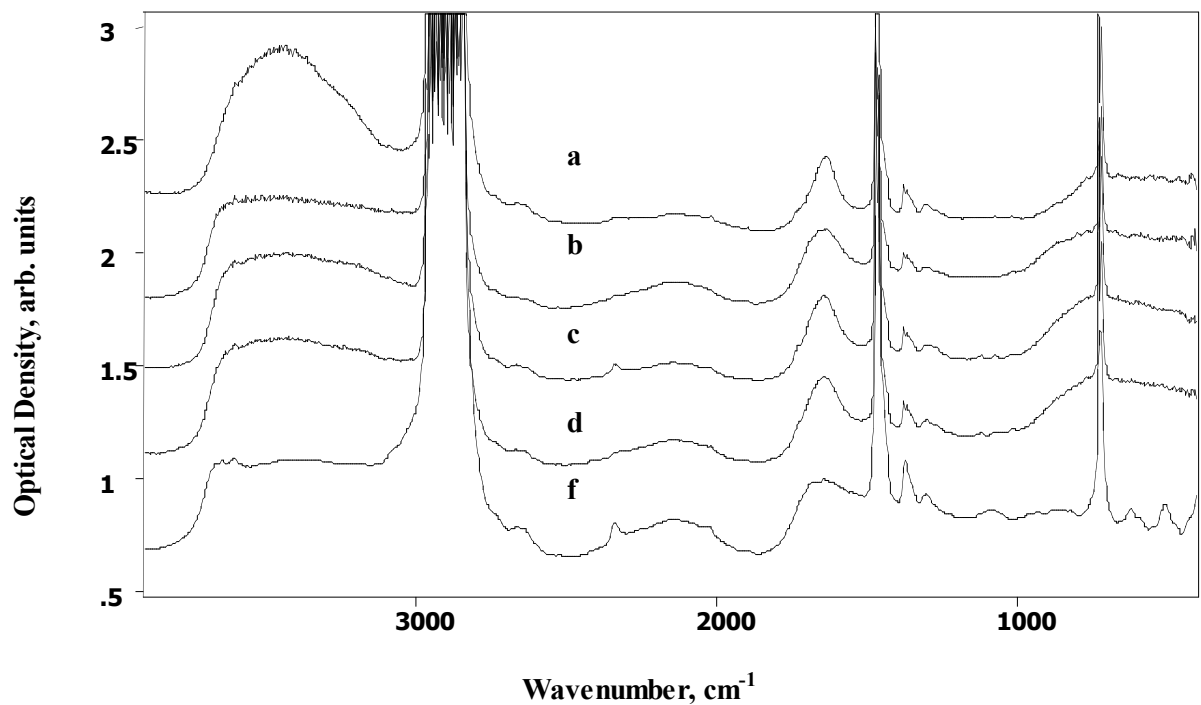

Fig.1. FTIR spectra of mineral water: a – distillate, b – Christinen, c – Bad Harzburger, d – Aqua Montana, f – Slavyanovskaya.

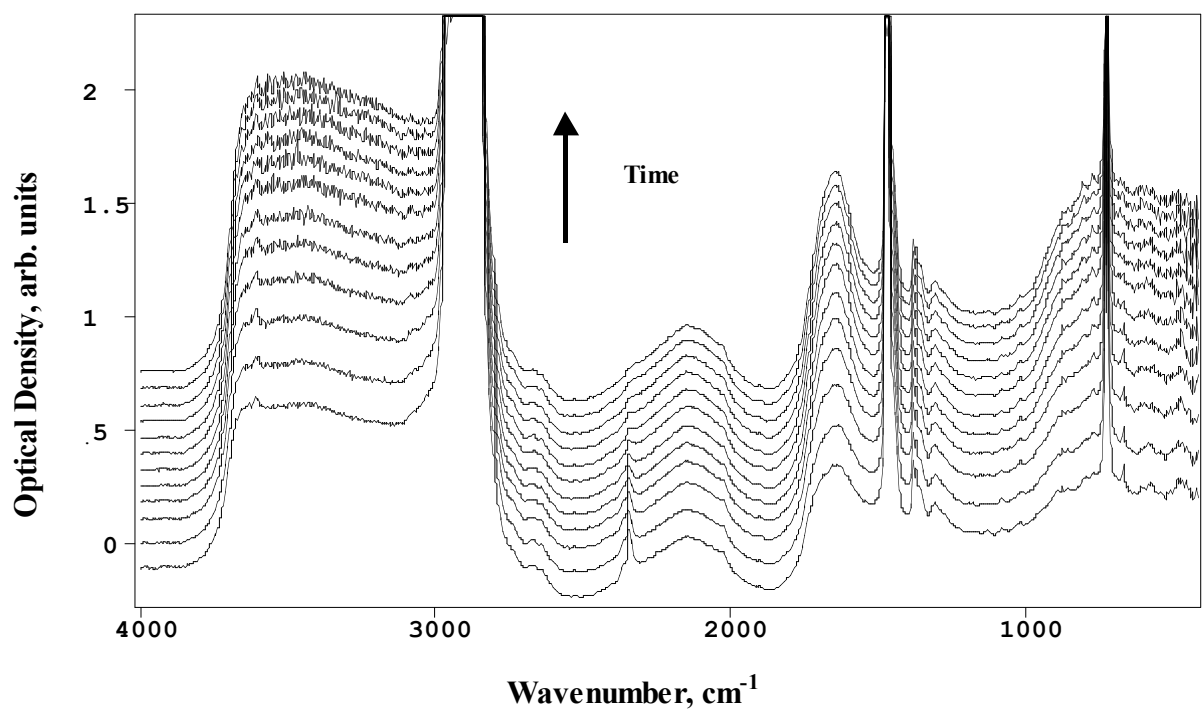

Fig.2. FTIR spectra of Christinen mineral water with time after getting from bottle. Time steps are following time scale in Fig.3.

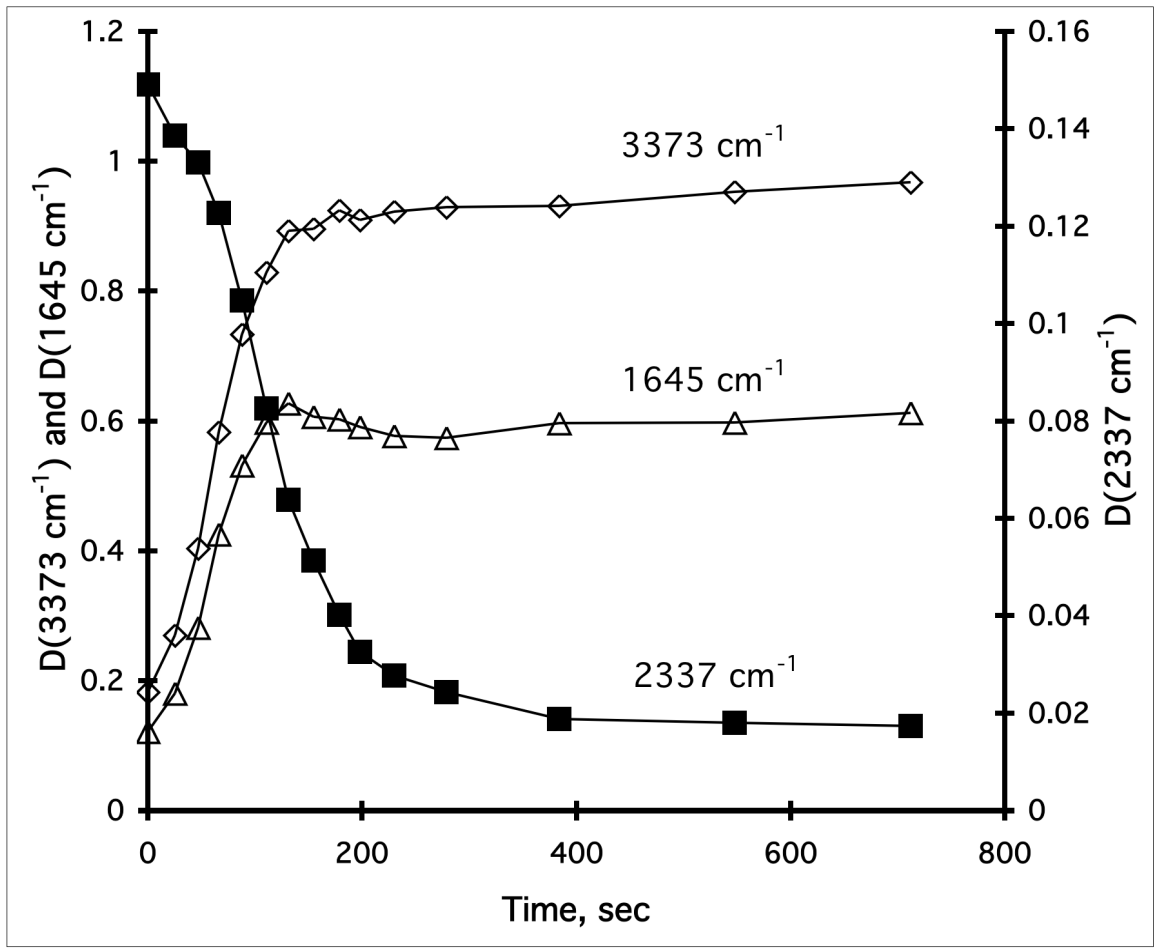

Fig.3. Absorbance of stretch (rhombus, 3373 cm$^{-1}$) and deformation (triangle, 1645 cm$^{-1}$) vibration line of water molecules and absorbance of carbon dioxide vibration line (cubic, 2337 cm$^{-1}$) in FTIR spectra of Aqua Montana mineral water with time after getting from bottle.

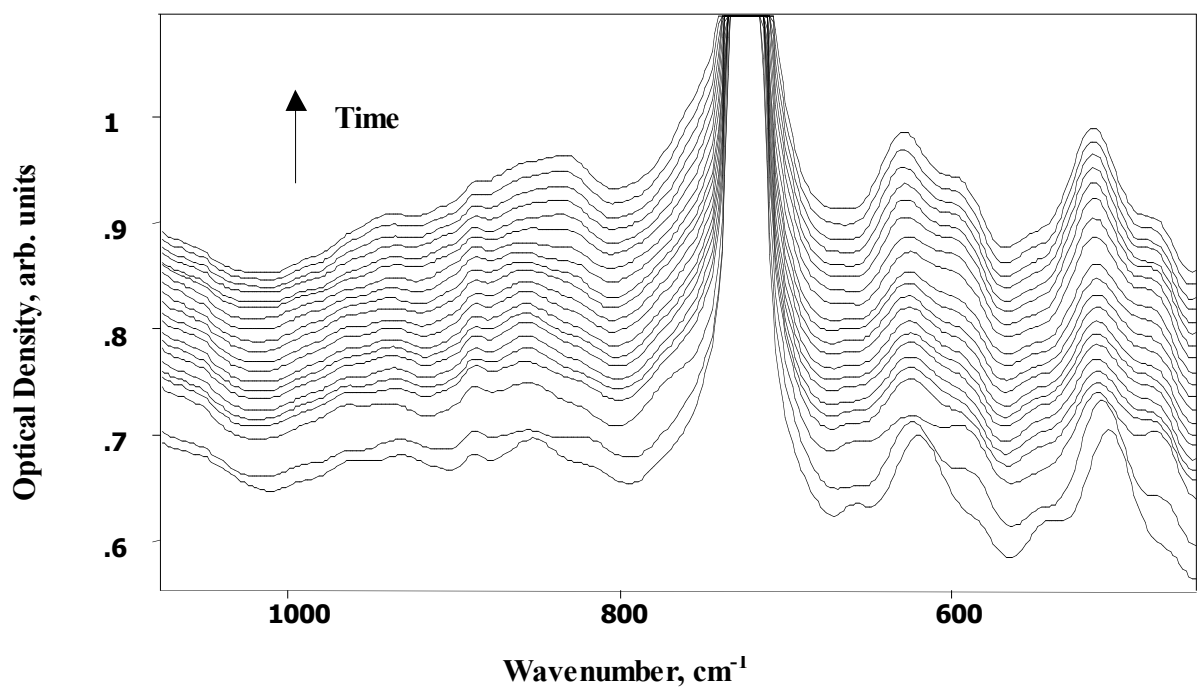

Fig.4a. FTIR spectra of Slavyanovskaya mineral water with time after getting from container. Time steps are following time scale in Fig.5.

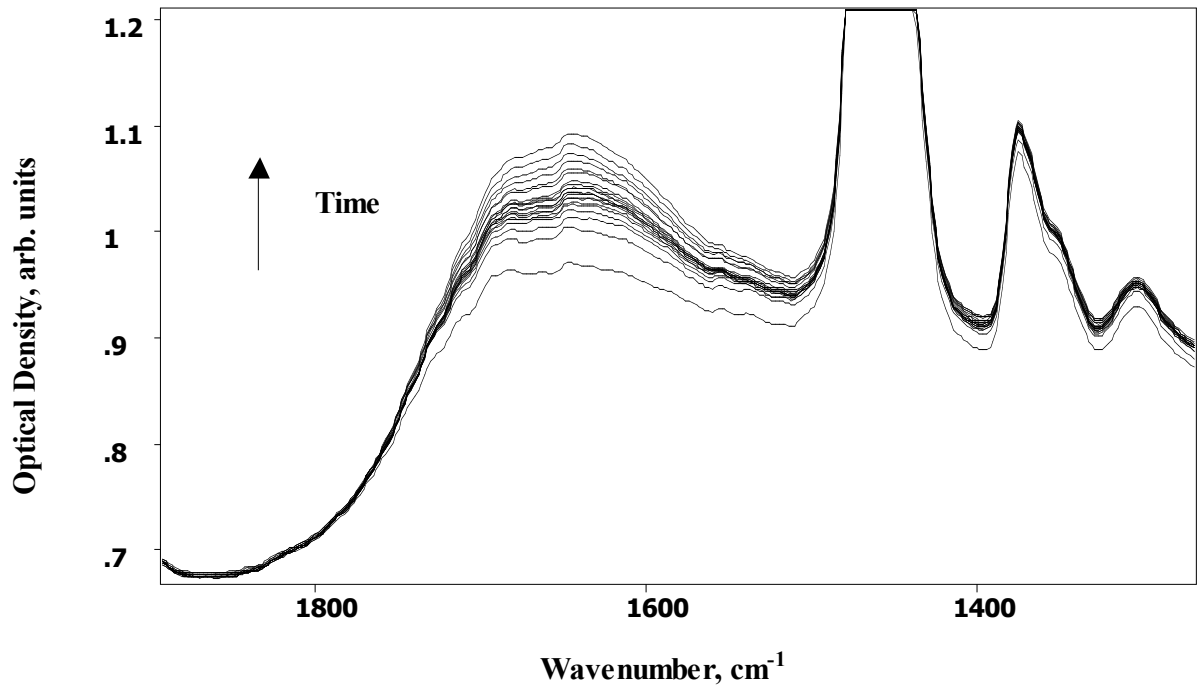

Fig.4b. FTIR spectra of Slavyanovskaya mineral water with time after getting from container. Time steps are following time scale in Fig.5.

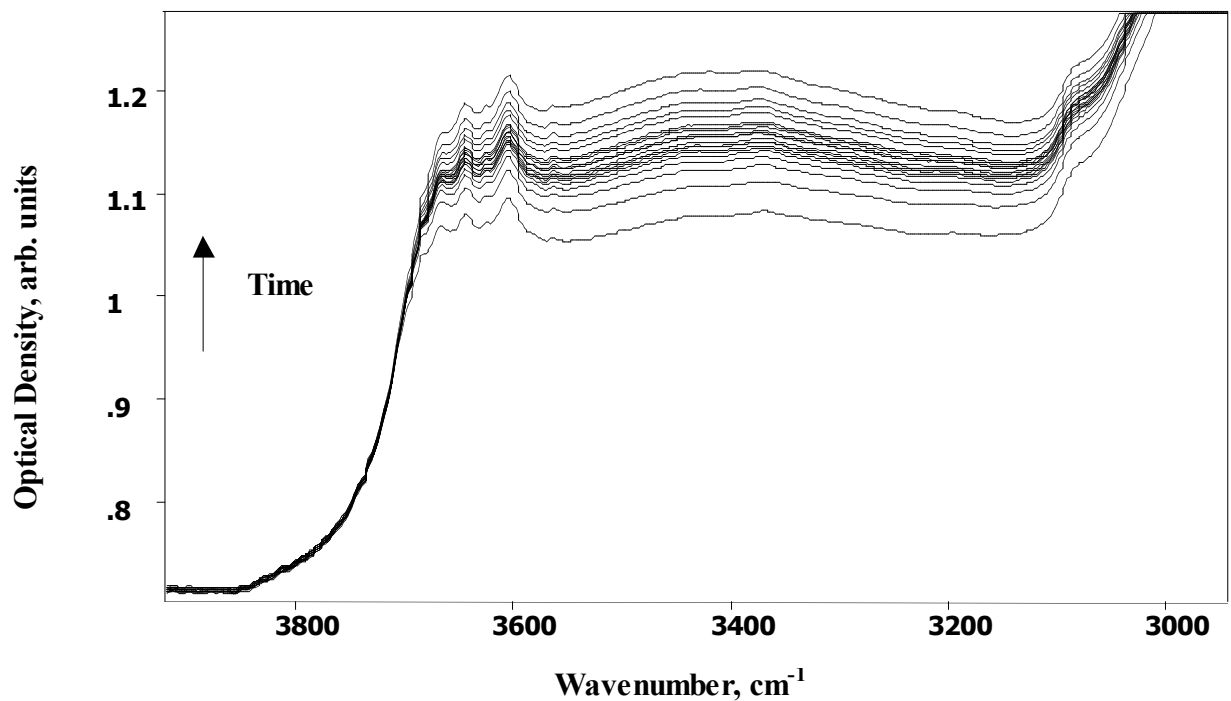

Fig.4c. FTIR spectra of Slavyanovskaya mineral water with time after getting from container. Time steps are following time scale in Fig.5.

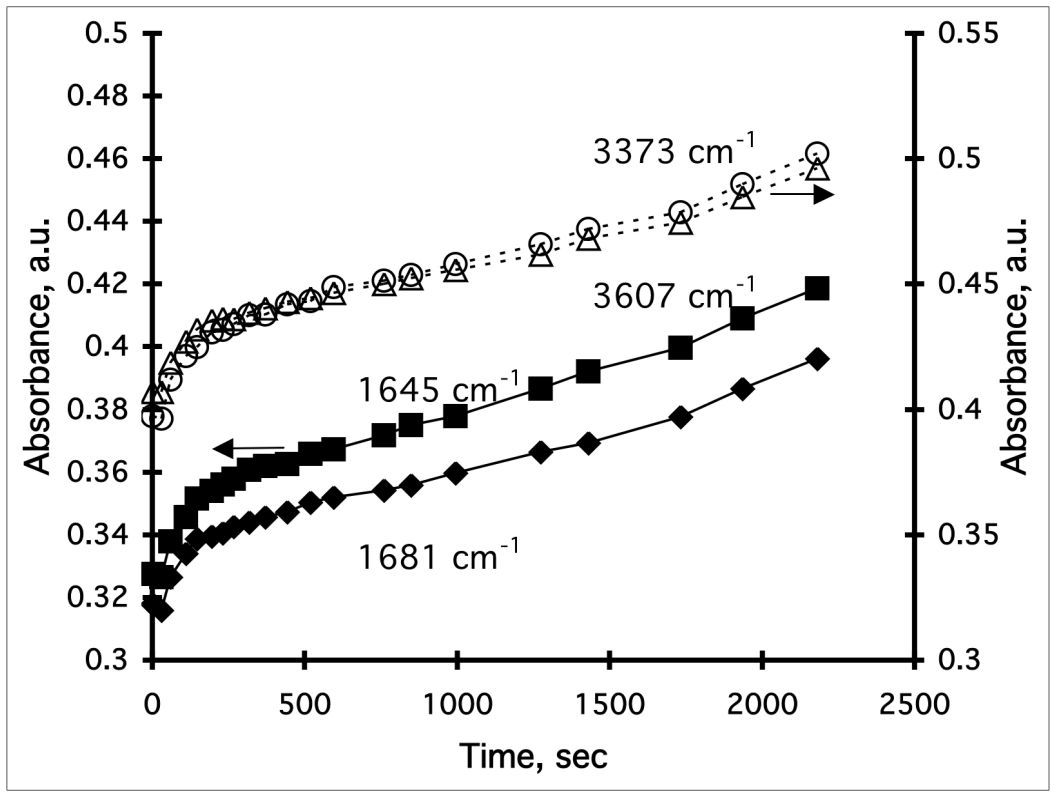

Fig.5. Absorbance of stretch and deformation vibration lines in FTIR spectra of active Slavyanovskaya mineral water with time after extracting from container. Notification of curves: triangle - 3607 cm$^{-1}$ stretching vibration, circle - 3373 cm$^{-1}$ stretching vibration, rhombus - 1681 cm$^{-1}$ deformation vibrations, cubic - 1645 cm$^{-1}$ deformation vibrations.

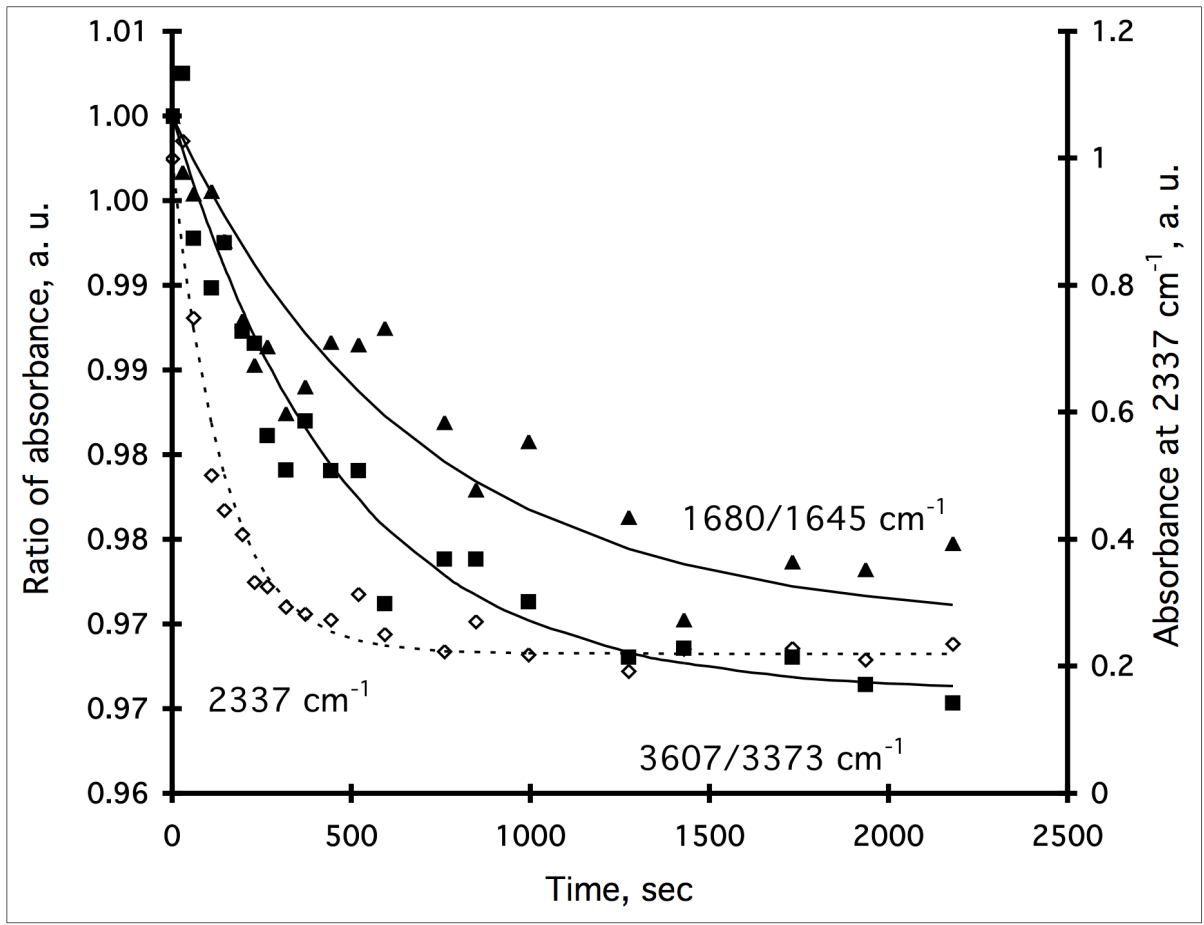

Fig. 6. Dependence of absorbance ratio on time scale in spectra of Slavyanovskaya mineral water: cubic - stretch vibrations (3607/3373 cm$^{-1}$), triangle - deformation vibrations (1680/1645 cm$^{-1}$). Rhombus (light) shows absorbance of carbon dioxide vibrations.

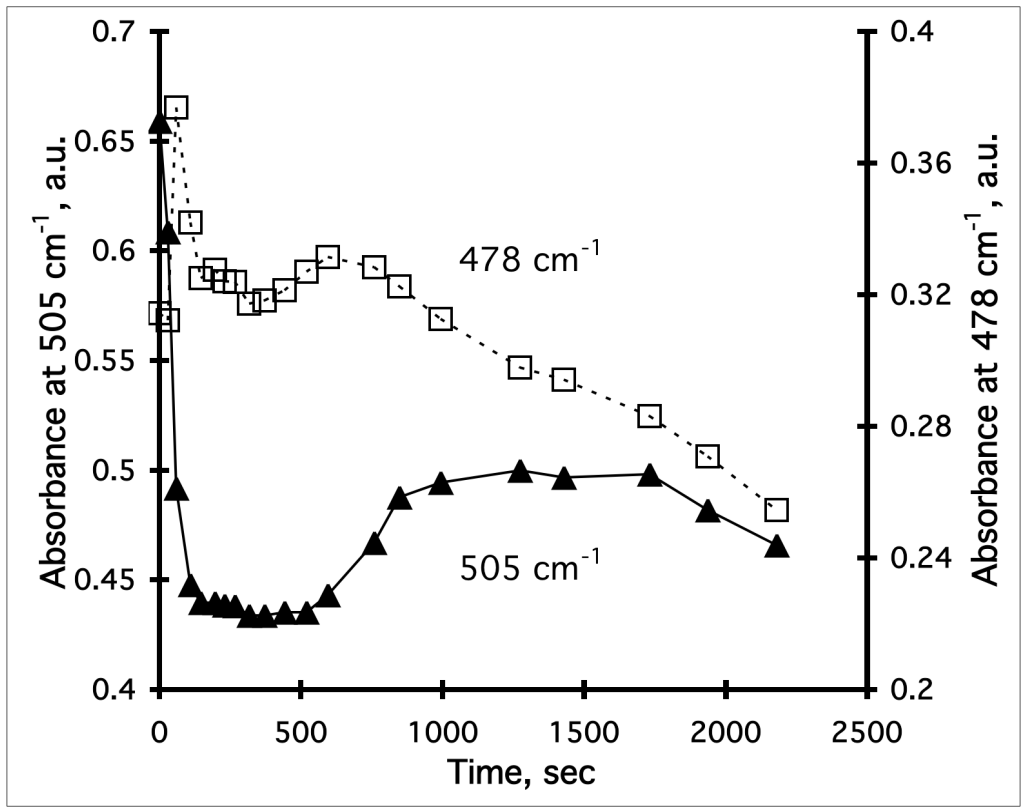

Fig.7. Absorbance of intermolecular vibration lines of FTIR spectra of Slavyanovskaya mineral water with time after getting from container. Notification of curves: triangle - 505 cm$^{-1}$, left scale; cubic - 478 cm$^{-1}$, right scale.